\documentstyle[12pt]{article}
\def\one{1\hskip-.37em 1}     
\def\l{\lambda}
\def\D{{\cal D}}
\def\H{{\cal H}}
\def\E{{\rm E}\hskip-.55em{\rm I}}
\def\ir{{\rm I}\hskip-.2em{\rm R}}
\def\half{\textstyle{\frac{1}{2}}}

\def\ra{\rightarrow}
\def\tint{{\textstyle{\int}}}
\def\d{\partial}
\def\o{\overline}
\def\b{\begin{eqnarray*}}     
\def\e{\end{eqnarray*}}       
\def\bn{\begin{eqnarray}}     
\def\en{\end{eqnarray}}       
\def\<{\langle}
\def\>{\rangle}
\def\{{\lbrace}
\def\}{\rbrace}
\title{Coherent State Path Integrals for \\Systems with 
Constraints\footnote{To appear in the proceedings of the International 
Seminar ``Path Integrals: Theory \& Applications'' and 5th International 
Conference on Path Integrals from meV to MeV, Dubna, Russia, May 27-31, 
1996\vskip.3cm \noindent BUTP-96/18}}
\author{\large John R. Klauder\\[3mm]
\em Departments of Physics and Mathematics\\
\em University of Florida\\
\em Gainesville, Fl  32611\\
 and\\
\em Institut f\"ur Theoretische Physik\\
\em Universit\"at Bern\\
\em Bern, CH-3012}
\date{\today}
\begin{document}
\maketitle
\begin{abstract}
We outline the principal results of a recent examination of the quantization 
of systems with first- and second-class constraints from the point of view 
of coherent-state 
phase-space path integration. Two examples serve to illustrate the 
procedures.
\end{abstract}
\section{Introduction}
\subsubsection*{Classical backround}
The quantization of systems with constraints is of considerable importance in a variety of applications. 
Let $\{p_j,q^j\}$, $1\leq j\leq J$, denote a set of dynamical variables, $\{\l^a\}$, $1\leq a\leq A\leq 2J$, a set of Lagrange multipliers, and $\{\phi_a(p,q)\}$ a set of constraints. Then the dynamics of a constrained system may be summarized in the form of an action principle by means of the classical action (summation implied)
  \bn I=\tint[p_j{\dot q}^j-H(p,q)-\l^a\phi_a(p,q)]\,dt\;.  \en
 The resultant equations that arise from the action read
  \bn  &&{\dot q}^j=\frac{\d H(p,q)}{\d p_j}+\l^a\frac{\d\phi_a(p,q)}{\d p_j}\equiv\{q^j,H\}+\l^a\{q^j,\phi_a\}\;,\nonumber\\
  &&{\dot p}_j=-\frac{\d H(p,q)}{\d q^j}-\l^a\frac{\d\phi_a(p,q)}{\d q^j}\equiv\{p_j,H\}+\l^a\{p_j,\phi_a\}\;,\nonumber\\
  &&\phi_a(p,q)=0\;,  \en
where $\{\cdot,\cdot\}$ denotes the Poisson bracket. The set of conditions $\{\phi_a(p,q)=0\}$ define the {\it constraint hypersurface}. If the constraints satisfy 
\bn &&\{\phi_a(p,q),\phi_b(p,q)\}=c_{ab}^{\;\;\;\;c}\,\phi_c(p,q)\;,\\
  &&\{\phi_a(p,q),H(p,q)\}=h_a^{\;\;b}\,\phi_b(p,q)\;,  \en
then we are dealing with a system of first-class constraints. If the coefficients $c_{ab}^{\;\;\;\;c}$ and $h_a^{\;\;b}$ are constants, then it is a closed system of first-class constraints; if they are suitable functions of the variables $p,q$, then it is called an open first-class constraint system. If one or both of the conditions in (3) or (4) fails, then the constraints are said to be second class. 

For first-class constraints it is sufficient to impose the constraints at the initial time inasmuch as the equations of motion will ensure that the constraints are fulfilled at all future times. Such an initial imposition of the constraints is called an {\it initial value equation}. Furthermore, the Lagrange multipliers are not determined by the equations of motion; rather they must be specified (a choice of ``gauge'') in order for a solution of the dynamical equations to be given. For second-class constraints, on the other hand, the Lagrange multipliers are determined by the equations of motion in such a way that the constraints are satisfied for all time. 

In the remainder of this section we review standard quantization procedures for systems with closed first-class constraints, both of the operator and path integral variety, pointing out some problems in each approach. In the following two sections we develop our coherent state approach, first for closed first-class constraints, and second for general constraints, i.e., open first-class constraints as well as second-class constraints \cite{kla}.

Due to space limitations, we are only able to offer here a few examples; the reader may consult \cite{kla} for a discussion of additional examples.
\subsubsection*{Standard operator quantization}
For a system of closed first-class constraints we assume (with $\hbar=1$) that 
\bn &&[\Phi_a(P,Q),\Phi_b(P,Q)]=ic_{ab}^{\;\;\;\;c}\,\Phi_c(P,Q)\;,\\
  &&[\Phi_a(P,Q),{\cal H}(P,Q)]=ih_a^{\;\;b}\,\Phi_b(P,Q)\;,  \en
where $\Phi_a$ and $\H$ denote self-adjoint constraint and Hamiltonian operators, respectively. Following Dirac \cite{dir}, we adopt the quantization prescription given by
  \bn i{\dot W}(P,Q)=[W(P,Q),{\cal H}(P,Q)]  \en
where $W$ denotes any function of the kinematical operators $\{Q^j\}$ and $\{P_j\}$ which are taken as a self-adjoint, irreducible representation of the commutation rules $[Q^j,P_k]=i\delta^j_k\one$, with all other commutators vanishing. The equations of motion hold for all time $t$, say $0<t<T$. On the other hand, the conditions
  \bn  \Phi_a(P,Q)|\psi\>_{\rm phys}=0  \en
to select the physical Hilbert space are imposed only at time $t=0$ as the analog of the initial value equation; the quantum equations of motion ensure that the constraint conditions are fulfilled for all time.

The procedure of Dirac has potential difficulties if zero lies in the continuous spectrum of the constraint operators for in that case there are no normalizable solutions of the constraint condition. We face the same problem, of course, and our resolution is discussed in detail in Ref.~\cite{kla}.
\subsubsection*{Standard path integral quantization}
Faddeev \cite{fad} has given a path integral formulation in the case of closed first-class constraint systems as follows. The formal path integral
 \bn &&\int\exp\{i\tint_0^T[p_j{\dot q}^j-H(p,q)-\l^a\phi_a(p,q)]\,dt\}\,\D p\,\D q\,\D\l \nonumber\\
  &&\hskip1.5cm=\int\exp\{i\tint_0^T[p_j{\dot q}^j-H(p,q)]\,dt\}\,\delta\{\phi(p,q)\}\,\D p\,\D q \en
may well encounter divergences in the remaining integrals. Therefore, subsidiary conditions in the form $\chi^a(p,q)=0$, $1\leq a\leq A$, are imposed picking out (ideally) one gauge equivalent point per gauge orbit, and in addition a  factor (in the form of a determinant) is introduced to formally preserve canonical covariance. The result is the path integral 
 \bn  \int\exp\{i\tint_0^T[p_j{\dot q}^j-H(p,q)]\,dt\}\,\delta\{\chi(p,q)\} \det(\{\chi^a,\phi_b\})\delta\{\phi(p,q)\}\,\D p\,\D q\,.  \en
This result may also be expressed as
 \bn \int\exp\{i\tint_0^T[p^*_j{\dot q}^{*j}-H^*(p^*,q^*)]\,dt\}\,\D p^*\,\D q^*\;,  \en
namely, as a path integral
over a reduced phase space in which the $\delta$-functionals have been used to eliminate $2A$ integration variables. 

The final expression generally involves curvilinear phase-space coordinates for which the definition of the path integral is typically ill defined. Additionally, in the form (10), the Faddeev-Popov determinant often suffers from ambiguities connected with inadmissible gauge fixing conditions \cite{gri}. Thus this widely used prescription is not without its difficulties.
\section{Coherent State Path Integral}
Canonical coherent states may be defined by the relation
 \bn  |p,q\>\equiv e^{-iq^jP_j}\,e^{ip_jQ^j}\,|0\>\;,  \en
where $|0\>$ traditionally denotes a normalized, unit frequency, harmonic oscillator ground state. Here $\{Q^j\}$ and $\{P_j\}$, $1\leq j\leq J$, denote an irreducible set of self-adjoint operators satisfying the Heisenberg commutation relations. The coherent states admit a resolution of unity in the form
 \bn  \one=\tint\,|p,q\>\<p,q|\,d\mu(p,q)\;,\hskip1.5cm d\mu(p,q)\equiv d^J\!p\,d^J\!q/(2\pi)^J\;,  \en
where the integration is over $\ir^{2J}$ and this integration domain and the form of the measure are unique. For a general operator ${\cal H}(P,Q)$ we introduce the upper symbol
   \bn  H(p,q)\equiv\<p,q|{\cal H}(P,Q)|p,q\>=\<p,q|:H(P,Q):|p,q\> \;  \en
which is related to the normal-ordered form as shown. If $\cal H$ denotes the quantum Hamiltonian, then we shall adopt $H(p,q)$ as the classical Hamiltonian. We also note that an important one-form is given by
$i\<p,q|d|p,q\>=p_j\,dq^j$. 

Using these quantities, the coherent state path integral for the  time-dependent Hamiltonian ${\cal H}(P,Q)+\l^a(t)\Phi_a(P,Q)$ is readily given by
\bn  &&\<p'',q''|{\sf T}e^{-i\tint_0^T[{\cal H}(P,Q)+\l^a(t)\Phi_a(P,Q)]\,dt}|p',q'\>\nonumber\\
&&\hskip.6cm=\lim_{\epsilon\ra0}\int\prod_{l=0}^N\<p_{l+1},q_{l+1}|e^{-i\epsilon({\cal H}+\l^a_l\Phi_a)}\,|p_l,q_l\>\prod_{l=1}^N\,d\mu(p_l,q_l)\nonumber\\
&&\hskip.6cm=
 \int\exp\{i\tint[i\<p,q|(d/dt)|p,q\>-\<p,q|{\cal H}+\l^a\Phi_a|p,q\>]\,dt\}\,\D\mu(p,q)\nonumber\\
&&\hskip.6cm= {\cal M}\int\exp\{i\tint[p_j{\dot q}^j-H(p,q)-\l^a\phi_a(p,q)]\,dt\}\,\D p\,\D q \;.  \en
In the second line we have set $p_{N+1},q_{N+1}=p'',q''$ and $p_0,q_0=p',q'$, and repeatedly inserted the resolution of unity; in the third and fourth lines we have formally interchanged the continuum limit and the integrations, and written for the integrand the form it assumes for continuous and differential paths ($\cal M$ denotes a formal normalization constant). The result evidently depends on the chosen form of the functions $\{\l^a(t)\}$. 
\subsubsection*{Enforcing the quantum constraints}
Let us next introduce the quantum analog of the initial value equation. For simplicity we assume that the constraint operators form a compact group; the case of a noncompact group is dealt with in \cite{kla}. In that case 
  \bn \E\,\equiv\tint e^{-i\xi^a\Phi_a(P,Q)}\,\delta\xi  \en
defines a {\it projection operator} onto the subspace for which $\Phi_a=0$ provided that $\delta\xi$ denotes the normalized, $\tint\delta\xi=1$, group invariant measure. Based on (5) and (6) it follows that
  \bn && \hskip1.5cm e^{-i\tau^a\Phi_a}\E\,=\E\,\;,  \\
      && e^{-i{\cal H}T}\E\,=\E\,e^{-i{\cal H}T}\E\,=\E\,e^{-i(\E\,{\cal H}\E\,)T}\E\,\;.  \en
We now project the propagator (15) onto the quantum constraint subspace 
which leads to the following set of relations
 \bn &&\hskip-1cm\int\<p'',q''|{\sf T}e^{-i\tint[{\cal H}+\l^a(t)\Phi_a]\,dt}\,|{\o p}',{\o q}'\>\<{\o p}',{\o q}'|\E\,|p',q'\>\,d\mu({\o p}',{\o q}')\nonumber\\
&&=\<p'',q''|{\sf T}e^{-i\tint[{\cal H}+\l^a(t)\Phi_a]\,dt}\,\E\,|p',q'\>\nonumber\\
&&=\lim\,\<p'',q''|[\prod^{\leftarrow}_l(e^{-i\epsilon{\cal H}}e^{-i\epsilon\l^a_l\Phi_a})]\,\E\,|p',q'\>\nonumber\\
&&=\<p'',q''|e^{-iT{\cal H}}e^{-i\tau^a\Phi_a}\,\E\,|p',q'\>\nonumber\\
&&=\<p'',q''|e^{-iT{\cal H}}\,\E\,|p',q'\>\;,  \en
where $\tau^a$ incorporates the functions $\l^a$ as well as the structure parameters $c_{ab}^{\;\;\;\;c}$ and $h_a^{\;\;b}$.
Alternatively, this expression has the formal path integral representation
 \bn \int\exp\{i\tint[p_j{\dot q}^j-H(p,q)-\l^a\phi_a(p,q)]\,dt-i\xi^a\phi_a(p',q')\}\,\D\mu(p,q)\,\delta\xi\;.  \en
On comparing (19) and (20) we observe that {\it after projection onto the quantum constraint subspace the propagator is entirely independent of the choice of the Lagrange mutiplier functions. In other words, the projected propagator is gauge invariant.} 

We may also express the physical (projected) propagator in a more general form, namely,
\bn &&\hskip-1cm\int\exp\{i\tint[p_j{\dot q}^j-H(p,q)-\l^a\phi_a(p,q)]\,dt\}\,\D\mu(p,q)\,\D C(\l)\nonumber\\
&&\hskip.1cm=\<p'',q''|e^{-iT{\cal H}}\,\E\,|p',q'\>  \en 
provided that $\tint\D C(\l)=1$ and that such an average over the functions $\{\l^a\}$ introduces (at least) one factor $\E\,$. 

\section{Application to General Constraints}
\subsubsection*{Classical considerations}
When dealing with a general constraint situation it will typically happen that the self-consistency of the equations of motion will determine some or all of the Lagrange multipliers in order for the system to remain on the classical constraint hypersurface. For example, if the Hamiltonian attempts to force points initially lying on the constraint hypersurface to leave that hypersurface, then the Lagrange multipliers must supply the necessary forces for the system to remain on the constraint hypersurface. 
\subsubsection*{Quantum considerations}
As in the previous section we let $\E\,$ denote the projection operator onto the quantum constraint subspace. Motivated by the classical comments given above we consider the quantity
\bn \lim\,\<p'',q''|\E\,e^{-i\epsilon{\cal H}}\E\,e^{-i\epsilon{\cal H}}\cdots\E\,e^{-i\epsilon{\cal H}}\E\,|p',q'\> \en
where the limit, as usual, is for $\epsilon\ra0$. The physics behind this expression is as follows. Reading from right to left we first impose the quantum initial value equation, and then propagate for a small amount of time ($\epsilon$). Next we recognize that the system may have left the quantum constraint subspace, and so we project it back onto that subspace, and so on over and over. In the limit that $\epsilon\ra0$ the system remains within the quantum constraint subspace and (22) actually leads to
 \bn \<p'',q''|\E\,e^{-iT(\E\;{\cal H}\E\;)}\E\,|p',q'\>\;, \en
which clearly illustrates temporal evolution entirely within the quantum constraint subspace. If we assume that $\E\,{\cal H}\E\,$ is a self-adjoint operator, then we conclude that (23) describes a unitary time evolution within the quantum constraint subspace. 

The expression (22) may be developed in two additional ways. First, we repeatedly insert the resolution of unity in such a way that (22) becomes
 \bn \lim\,\int\prod_{l=0}^N\<p_{l+1},q_{l+1}|\E\,e^{-i\epsilon{\cal H}}\E\,|p_l,q_l\>\prod_{l=1}^Nd\mu(p_l,q_l)\;. \en
We wish to turn this expression into a formal path integral, but the procedure used previously relied on the use of unit vectors, and the vectors $\E\,|p,q\>$ are generally not unit vectors. Thus let us rescale the factors in the integrand introducing \bn  |p,q\>\!\>\equiv\E\,|p,q\>/\|\E\,|p,q\>\| 
 \en which are unit vectors. If we let $M''=\|\E\,|p'',q''\>\|$, $M'=\|\E\,|p',q'\>\|$, and observe that $\|\E\,|p,q\>\|^2=\<p,q|\E\,|p,q\>$, it follows that (24) may be rewritten as
 \bn M''M'\lim\,\int\prod_{l=0}^N\<\!\<p_{l+1},q_{l+1}|e^{-i\epsilon{\cal H}}|p_l,q_l\>\!\>\prod_{l=1}^N\<p_l,q_l|\E\,|p_l,q_l\>\,d\mu(p_l,q_l)\;. \en
This expression is represented by the formal path integral
\bn M''M'\int\exp\{i\tint[i\<\!\<p,q|(d/dt)|p,q\>\!\>-\<\!\<p,q|{\cal H}|p,q\>\!\>]\,dt\}\,\D_E\mu(p,q)\;,  \en
where the new formal measure for the path integral is defined in an evident fashion from its lattice prescription. We can also reexpress this formal path integral in terms of the original bra and ket vectors in the form 
 \bn &&\hskip-1cm M''M'\int\exp\{i\tint[i\<p,q|\E\,(d/dt)\E\,|p,q\>/\<p,q|\E\,|p,q\>\nonumber\\
    &&\hskip2.3cm-\<p,q|\E\,{\cal H}\E\,|p,q\>/\<p,q|\E\,|p,q\>]\,dt\}\,\D_E\mu(p,q)\;.\en
This last relation concludes our second route of calculation beginning with (22).

The third relation we wish to derive uses an integral representation for the projection operator $\E\,$ generally given by
  \bn \E\,=\tint e^{-i\xi^a\Phi_a(P,Q)}\,f(\xi)\,\delta\xi  \en
for a suitable function $f$. Thus we rewrite (22) in the form
\bn &&\hskip-1.5cm\lim\int\<p'',q''|e^{-i\epsilon\l^a_{N}\Phi_a}e^{-i\epsilon{\cal H}}e^{-i\epsilon\l^a_{N-1}\Phi_a}e^{-i\epsilon{\cal H}}\cdots e^{-i\epsilon\l^a_1\Phi_a}e^{-i\epsilon{\cal H}}e^{-i\epsilon\l^a_0\Phi_a}|p',q'\>\nonumber\\
&&\hskip2cm \times\,f(\epsilon\l_{N})\cdots f(\epsilon\l_0)\,\delta\epsilon\l_{N}\cdots\delta\epsilon\l_0\;.  \en
Next we insert the coherent-state resolution of unity at appropriate places to find that (30) may also be given by
 \bn &&\hskip-1.5cm\lim\int\<p_{N+1},q_{N+1}|e^{-i\epsilon\l^a_{N}\Phi_a}|p_N,q_N\>\prod_{l=0}^{N-1}\<p_{l+1},q_{l+1}|e^{-i\epsilon{\cal H}}e^{-i\epsilon\l^a_{l}\Phi_a}|p_l,q_l\>\nonumber\\   &&\times [\prod_{l=1}^{N}d\mu(p_l,q_l)\,f(\epsilon\l_l)\,\delta\epsilon\l_l]\,f(\epsilon\l_0)\,\delta\epsilon\l_0\;.  \en
Following the normal pattern, this last expression may readily be turned into a formal coherent-state path integral given by
\bn \int\exp\{i\tint[p_j{\dot q}^j-H(p,q)-\l^a\phi_a(p,q)]\,dt\}\,\D\mu(p,q)\D E(\l)\;,  \en
where $E(\l)$ is a measure designed so as to insert the projection operator $\E\,$ at every time slice. Unlike the case of the first-class constraints, we observe that the measure on the Lagrange multipliers is fixed. This usage of the Lagrange multipliers to ensure that the quantum system remains within the quantum constraint subspace is similar to their usage in the classical theory to ensure that the system remains on the classical constraint hypersurface. Thus it is not surprising that a fixed integration measure emerges for the Lagrange multipliers. On the other hand, it is also possible to use the measure $E(\l)$ in the case of closed {\it first}-class constraints as well; this would be just one of the acceptable choices for the measure $C(\l)$ designed to put at least one projection operator $\E\,$ into the propagator.

In summary, we have established the equality of the three expressions
\bn &&\hskip-1cm\<p'',q''|\E\;e^{-iT(\E\;{\cal H}\E\;)}\E\,|p',q'\>\nonumber\\
&&=M''M'\int\exp\{i\tint[i\<p,q|\E\,(d/dt)\E\,|p,q\>/\<p,q|\E\,|p,q\>\nonumber\\
    &&\hskip2.3cm-\<p,q|\E\,{\cal H}\E\,|p,q\>/\<p,q|\E\,|p,q\>]\,dt\}\,\D_E\mu(p,q)\nonumber\\
&&= \int\exp\{i\tint[p_j{\dot q}^j-H(p,q)-\l^a\phi_a(p,q)]\,dt\}\,\D\mu(p,q)\D E(\l)\;.  \en
This concludes our derivation of path integral formulas for general constraints. Observe that we have not introduced any $\delta$-functionals, nor, in the middle expression, reduced the number of integration variables or the limits of integration in any way even though in that expression the integral over the Lagrange multipliers has been effected. 
\section{Examples}
\subsubsection*{First-class constraint}
Consider the system with two degrees of freedom, a single constraint, and a vanishing Hamiltonian characterized by the action
 \bn I=\tint[\half(p_1{\dot q}_1-q_1{\dot p}_1+p_2{\dot q}_2-q_2{\dot p}_2)-\l(q_2p_1-p_2q_1)]\,dt\;,  \en
where for notational convenience we have lowered the index on the $q$ variables.
Note that we have chosen a different form for the kinematic part of the action which amounts to a change of phase for the coherent states. It follows in this case that
 \bn &&{\cal M}\int\exp\{i\tint[\half(p_1{\dot q}_1-q_1{\dot p}_1+p_2{\dot q}_2-q_2{\dot p}_2)-\l(q_2p_1-p_2q_1)]\,dt\}\,\D p\,\D q\,\D C(\l)\nonumber\\
&&\hskip2cm=\<p'',q''|\E\,|p',q'\>\;,  \en
where
  \bn  \E\,=(2\pi)^{-1}\int_0^{2\pi}e^{-i\xi(Q_2P_1-P_2Q_1)}\,d\xi=\E\,(L_3=0)\;.  \en
Based on the fact that
 \bn \<p'',q''|p',q'\>=\exp(-\half|z''_1|^2-\half|z''_2|^2+z''^*_1z'_1+z''^*_2z'_2-\half|z'_1|^2-\half|z'_2|^2)\;,  \en
where $z'_1\equiv(q'_1+ip'_1)/\sqrt{2}$, etc., it is straightforward to show that
 \bn  &&\<p'',q''|\E\,|p',q'\>  =\exp(-\half|z''_1|^2-\half|z''_2|^2-\half|z'_1|^2-\half|z'_2|^2)\nonumber\\
 &&\hskip3.3cm\times I_0(\sqrt{(z''^{*2}_1+z''^{*2}_2)(z'^2_1+z'^2_2)}\,)\;,\en
with $I_0$ a standard Bessel function. We emphasize again that although the Hilbert space has been reduced by the introduction of $\E\,$, the reproducing kernel (38) leads to a reproducing kernel Hilbert space with an inner product having the same number of integration variables and domain of integration as in the unconstrained case.
\subsubsection*{Second-class constraint}
Consider the two degree of freedom system determined by
\bn  I=\tint[p{\dot q}+r{\dot s}-H(p,q,r,s)-\l_1r-\l_2s]\,dt\;, \en
where we have called the variables of the second degree of freedom $r,s$, and $H$ is not specified further. The coherent states satisfy $|p,q,r,s\>=|p,q\>\otimes|r,s\>$, which will be useful. We adopt (28) as our formal path integral in the present case, and choose \cite{kla}
  \bn   &&\E\,=\tint e^{-i(\xi_1R+\xi_2S)}\,e^{-(\xi_1^2+\xi_2^2)/4}\,d\xi_1d\xi_2/2\pi\nonumber\\  &&\hskip.4cm=|r=0,s=0\>\<r=0,s=0|\equiv|0_2\>\<0_2|  \en
which is a projection operator onto a coherent state for the second (constrained) degree of freedom only. With this choice it follows that
 \bn &&i\<p,q,r,s|\E\,(d/dt)\E\,|p,q,r,s\>/\<p,q,r,s|\E\,|p,q,r,s\>\nonumber\\
&&\hskip2cm=i\<p,q|(d/dt)|p,q\>-\Im(d/dt)\ln[\<0_2|r,s\>]\nonumber\\
&&\hskip2cm=p{\dot q}-\Im(d/dt)\ln[\<0_2|r,s\>]\;,  \en
and 
\bn &&\hskip-1cm\<p,q,r,s|\E\,\H(P,Q,R,S)\E\,|p,q,r,s\>/\<p,q,r,s|\E\,|p,q,r,s\>\nonumber\\  &&\hskip1cm=\<p,q,0,0|\H(P,Q,R,S)|p,q,0,0\>=H(p,q,0,0)\;. \en
Consequently, for this example, (28) becomes
 \bn {\cal M}\int\exp\{i\tint[p{\dot q}-H(p,q,0,0)]\,dt\}\,\D p\,\D q
\;\times\;\<r'',s''|0_2\>\<0_2|r',s'\>\;, \en
where we have used the fact that at every time slice
 \bn \tint\<r,s|\E\,|r,s\>\,dr\,ds/(2\pi)=\tint|\<0_2|r,s\>|^2\,dr\,ds/(2\pi)=1\;. \en

Observe that in this path integral quantization no variables have been eliminated nor has any domain of integration been reduced; moreover, the operators $R$ and $S$ have remained unchanged. The result in (43) is clearly a product of two distinct factors. The first factor describes the true dynamics as if we had solved for the classical constraints and substituted $r=0$ and $s=0$ in the classical action from the very beginning, while the second factor characterizes a one-dimensional Hilbert space for the second degree of freedom. Thus we can also drop the second factor completely as well as all the integrations over $r$ and $s$ and still retain the same physics. In this manner we recover the standard result without the use of Dirac brackets or having to eliminate the second-class constraints from the theory initially.

\section*{Acknowledgements}
J. Govaerts is thanked for his continued interest in this work.
 Some aspects of a coherent state quantization procedure that emphasizes projection operators for systems with closed first-class constraints have been anticipated by Shabanov \cite{sha}. Thanks are expressed to S. Shabanov for bringing this work to the author's attention. Projection operators for closed first-class constraints also appear in the text of Henneaux and Teitelboim \cite{hen}.

\end{document}